\documentclass{llncs}

\usepackage{color}
\usepackage{graphicx}
\usepackage{multirow}
\usepackage{subfig}
\usepackage{url}
\usepackage{hyperref}
\usepackage{listings,xcolor}
\begin{document}

\title{Software Enabled Security Architecture and Mechanisms for Securing 5G Network Services}
\author{%
Vijay Varadharajan\and Uday Tupakula \and Kallol Karmakar}%
\institute{
Advanced Cyber Security Research Centre, The University of Newcastle, Australia\\
\email{[vijay.varadharajan, uday.tupakula, KallolKrishna.Karmakar]@newcastle.edu.au} }

\maketitle
\begin{abstract}

The 5G network systems are evolving and have complex network infrastructures. There is a great deal of work in this area focused on meeting the stringent service requirements for the 5G networks. Within this context, security requirements play a critical role as 5G networks can support a range of services such as healthcare services, financial and critical infrastructures. 3GPP and ETSI have been developing security frameworks for 5G networks. Our work in 5G security has been focusing on the design of security architecture and mechanisms enabling dynamic establishment of secure and trusted end to end services as well as development of mechanisms to proactively detect and mitigate security attacks in virtualised network infrastructures. The focus of this paper is on the latter, namely the facilities and mechanisms, and the design of a security architecture providing facilities and mechanisms to detect and mitigate specific security attacks. We have developed and implemented a simplified version of the security architecture using Software Defined Networks (SDN) and Network Function Virtualisation (NFV) technologies. The specific security functions developed in this architecture can be directly integrated into the 5G core network facilities enhancing its security. We describe the design and implementation of the security architecture and demonstrate how it can efficiently mitigate specific types of attacks. 
\end{abstract}
\vspace{-0.2cm}
\section{Introduction}
\vspace{-0.3cm}

Future networked systems must provide services to a vast array of applications and devices with competing and even conflicting requirements while simultaneously allowing flexible deployment. Today’s networks, with their “one-size-fits-all” architectural approach, are unable to address the diverging performance requirements that vertical industries such as manufacturing, automotive, healthcare, energy, and media and entertainment impose in terms of latency, scalability, security, availability, and reliability. Two major trends in future networked systems include the fifth-generation mobile systems (5G) and software defined network architectures which are anticipated to open up innovation opportunities for satisfying such requirements in these applications. Increasingly such networks are highly heterogeneous with many different devices, from small sensors and IoT and network devices to many different clients and servers and peripherals. Such a complex heterogeneous environment supporting distributed applications and services with many network devices, supporting user and device mobility and dynamic variation in networks (e.g. due to failure of devices and network links) can be subjected to sophisticated security attacks.\\
The overall aim of our research work is to consider the choices involved in the design of security architecture and mechanisms for securing 5G services over multi-domain networks.
5G networks leverage softwarisation and virtualisation to achieve the service objectives on ﬂexibility, conﬁgurability, and scalability. In particular, key design concepts of 5G networks include network slicing (i.e., dedicating logical networks for isolated applications), mobile edge computing (MEC), NFV \cite{etsi2014002}, and SDN \cite{onf14}, \cite{nadeau13}. 
The security of 5G networks and their communication services are of vital importance as secure and reliable network services are a prerequisite and enabler for digital markets such as for Industry 4.0 and eHealth. However, there are several challenges arise, which are mainly due to the networks’ dynamic environment and the fact that the security requirements will be much more stringent than in previous network generations since the diverse network services from verticals can be mission critical. With 5G services, there is a need to take into account new types of trust relations between participating actors in the security design; whom is to be trusted, in which respect, and to what extent. Furthermore, the use of new technologies like network virtualisation (i.e., decoupling logical networks from networking hardware) and SDN will bring new trust issues; in this case trust between application owners and compute and storage resource providers. In both these cases, the trust relations will manifest themselves in hard security requirements to enforce required service level agreements and to protect information exchange between actors.\\
A cornerstone in developing secure systems is the design of a security architecture in terms of the different entities involved, and their relations and interactions. In the case of 5G, challenges in developing a security architecture arise not only from the technologies used in 5G but also from the use cases stemming from the new business environment offered by 5G. For instance, existing security architectures need to be designed for multi-tenancy operation (e.g. shared physical infrastructure used by different providers) and need to differentiate trust relations between the different tenants. Furthermore, the architecture needs to provide secure support for network virtualisation and network slicing (i.e., dedicating logical networks for isolated applications); these requirements are not satisfied by the existing security architectures. The overall aim of our security architecture is that it should enable dynamic establishment of secure and trusted end to end services over networked virtualised infrastructures as well as provide mechanisms to proactively detect and mitigate security attacks thereby enhancing network resiliency. \\
Given this overall context, the focus of this paper is on the latter, namely the facilities and mechanisms to detect and mitigate specific security attacks, and the design of a security architecture providing these facilities and mechanisms. We have developed and implemented this security architecture and experimented with specific security attacks demonstrating how they can be mitigated in a single 5G network operator. This is the main contribution of this paper. We are continually evolving this security architecture extending with additional security functionalities. \\
More specifically, the attacks and security services that are considered in this paper fall under three main categories:\\
(i) Facilities and mechanisms detecting and mitigating attacks from malicious user devices: mechanisms to ensure that only authorised user devices are able to access services provided by specific network slices as well as the ability to detect and prevent attacks from compromised but authorised user devices. \\
(ii) Facilities and mechanisms for monitoring infrastructure components in 5G networks: mechanisms for validating the state of these components, which in turn can be used in the provision of trusted services as well as used in attack detection, for instance, on network switches.\\  
(iii) Facilities and mechanisms for provisioning security-as-a-service in 5G networks: mechanisms enabling security-as-a-service to users and specific devices. For instance, for some IoT and legacy devices, which are resource-constrained or can lack capacity for hosting security functionalities.\\
\noindent We will then demonstrate the applicability of the proposed security architecture using an IoT based use case scenario. The selected use case represents a common scenario in 5G networks, as it is envisaged that they are likely to support a number of devices utilising their services managed by different actors.\\
\noindent The remainder of the paper is organized as follows: 
In Section \ref{sec:mot}, we present the use case scenarios for attacker model from which the specific security requirements that are to be addressed by the architecture in this paper are derived. In Section \ref{sec:sad}, the design of the security architecture and the security mechanisms for detecting and mitigating attacks are described. Then we discuss the use case scenarios in 5G networks demonstrating the applicability of the proposed security architecture. We describe the implementation of the developed architecture and the analysis of attacks in Section \ref{sec:imp}. Section \ref{sec:rel} discusses the relevant related works and 
 in Section \ref{sec:con} we draw some conclusions and outline further work. 

\vspace{-0.4cm}
\section{Motivating Use Case Scenario} \label{sec:mot}
\vspace{-0.3cm}
 \noindent Consider a scenario where each user can use multiple devices to access various services in different slices in 5G networks. For instance, a user uses his/her laptop (user equipment (UE1)) to carry out banking operations using a cloud service while at the same time his/her wearable device (user equipment (UE2)) is uploading health-related data to a server (S) in the cloud.  These devices are accessing services in two different network slices, namely the wearable device is using services in a healthcare slice whereas the mobile device is using banking services in a financial slice.  
 
\noindent When the gNodeB (in the 5G network) receives a new flow from UE1 to access financial slice, it needs to be analyzed to determine whether it is an authorised flow. This gives rise to the first requirement for the security architecture to have policies and mechanisms to determine whether a user is authorised to access specific slices, e.g. healthcare slice. These will help to detect attacks by unauthorised users on specific network slices \textbf{(Attack 1)}.

\noindent Now assume that the usage of a specific network function within the financial slice by UE1 exceeds the allowed limits. This gives rise to the second requirement for the security architecture to have policies and mechanisms for monitoring the usage of resources by authorised users. This will help to detect attacks by authorised users on specific network slices \textbf{(Attack 2)}.

\noindent The 5G network infrastructure components providing the various services in the 5G core and access networks (such as gNodeB) as well as in the Internet and cloud (such as routers and servers) might have been infected by malware leading to different attacks (such as DDoS flooding and botnets). This gives rise to the third requirement for the security architecture to have mechanisms to validate the state of these resources (both at boot and at runtime) \textbf{(Attack 3)}.

\noindent Furthermore, when accessing cloud and Internet services, the user can move between two different radio access networks (RANs) belonging to the same 5G network operator. This gives rise to the fourth requirement for the security architecture to ensure that the attacks are detected and mitigated closest to source of the origin of the attack \textbf{(Attack 4)}.

\noindent Our security architecture provides security services and mechanisms to counteract the attacks outlined in this motivating use case scenario.

\vspace{-0.4cm}

\section{Security Architecture: Design and Implementation} \label{sec:sad}
\vspace{-0.3cm}
\noindent Our network infrastructure consists of SDN enabled NFV deployment of network slices running on a network virtualisation infrastructure supporting 5G services. In such a framework, we have infrastructure providers who own and manage physical networks and its constituent resources. Tenant leases virtual resources from one or more infrastructure providers in the form of virtual machines and uses the virtual networks to provide network services to its users. The network slice is composed of a collection of virtual network functions (VNFs) that are composed to build the network services the tenant delivers to its end users. Each slice can have several VNFs, and each tenant can have several network slices and orchestrate its resources to satisfy the diverse requirements of the slices. \\
\noindent In this network infrastructure, there are different authorities performing different functions on different resources. The logical management authorities in our overall security architecture comprise the following: System Manager responsible for creating and managing tenants and their network slices; Network Slice Manager responsible for managing network slices comprising collections of VNFs for each tenant; VNF Manager responsible for creating and maintaining different VNFs; Virtual Infrastructure Manager (VIM) responsible for managing virtual cloud infrastructures. However, in our implementation described in this paper, we have combined these different authorities within a SDN Controller. 
\vspace{-0.3cm}
\subsection{Security Requirements}
\vspace{-0.2cm}
\noindent The security requirements for our architecture are based on the need to counteract the following attacks, from the motivating use case scenario above.\\
(i) Preventing access to slices by unauthorised devices (user equipment/user): The 5G networks  consists of several network slices with specific service requirements. For instance, higher network availability requirement for services dealing with emergency healthcare or control of critical infrastructures. Attackers can use their devices to generate attacks on multiple slices to impact adversely the performance of network functions in  services in multiple slices. For instance, a DDoS attack on one slice’s services can have adverse impact on another slice, if these two slices share some common VNFs. Hence the need to restrict the slice access to only authorised devices.\\
(ii) Devices authorised to access specific slices can generate attacks {\em within} these slices. For instance, attackers can compromise IoT devices that have access to specific slices and use these devices to generate attacks within the slices. Hence the need for the network slice manager to monitor and control the usage of slice resources by authorised devices and detect attacks within the slices. \\
(iii) The 5G infrastructure components themselves  can be vulnerable to different types of security attacks (e.g.infected by malware). This can lead to attacks originating from these infrastructure components, which can have more severe consequences than the attacks coming from the end devices and users. Often, network slice managers can be invoking VNFs on virtual infrastructures which may be deployed on physical devices in an inter-domain environment managed by a different provider. Hence there is need for techniques to validate the state of the infrastructure components before VNF deployment.\\
(iv) Compromise of a VNF can lead to unauthorised access to the virtualization infrastructure resources, enabling instantiation of new VNFs in insecure physical infrastructure components. Compromised VNFs can also lead to flooding attacks such as generation of a huge amount of logs or leaking of information.\\
(v) Each slice can have different service specific security requirements (depending on the services in the slice). There can be legacy or even resource constrained IoT devices in 5G networks, which do not have the capability to implement the required security functions. Such applications may wish to rely on security functions provided by the specific network slices. Or even  users may want to choose the security functions that are required for their applications (such as confidentiality, integrity etc.). There can also be situations where some legacy devices may require certain services; in these cases, it is critically important to ensure that these services are available only to these legacy devices and not to any other devices in the network. To cater for such requirements, there is a need for the security architecture to provision security-as-a-service in 5G networks. 
\vspace{-0.4cm}
\subsection{Security Architecture - Design}
{\bf Assumptions}:\\
(i)As indicated above, we assume there exists a single logical management authority in the form of SDN Controller per domain, which performs the various functions associated in the management of network slices, VNFs and virtual infrastructures.\\
(ii) We assume that SDN Controller and its various sub-components are not compromised and hence are  trusted (by other components in the network infrastructure and the users).\\
(iii) We assume that the network infrastructure components (such as physical and virtual switches, routers, gateways and servers) in the 5G network and the Internet are equipped with some trusted computing technology such as TPM chip~\cite{tpm03}. Most systems including servers, laptops and mobile devices that are manufactured today have such a TPM chip embedded in them. Hence it is reasonable to assume that the infrastructure components are equipped with TPM.

\vspace{-0.2cm}
\subsection{Security Management Application}
\begin{figure}
\centering
\includegraphics[scale=.5]{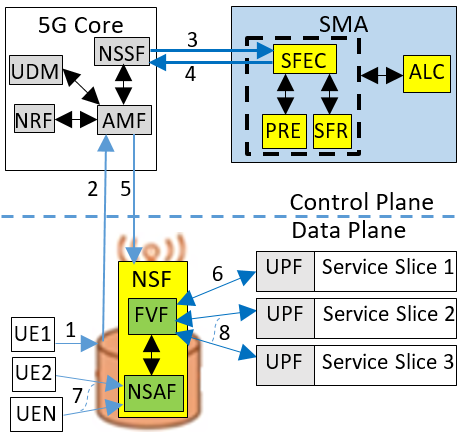}
\caption{Security Architecture for 5G Network}
\label{fig:secarc}
\end{figure}

\vspace{-0.2cm}
 \noindent A key component of our security architecture is the Security Management Application (SMA) hosted on the SDN Controller. Note as indicated in our assumptions, we consider SMA to be trusted and not compromised.  In this section, we describe the SMA and its components, namely Policy Repository and Engine (PRE), Security Function Repository (SFR), Security Function Enforcement Component (SFEC) and Activity Logs Component (ALC) (See Figure~\ref{fig:secarc}). \\
 \textbf{PRE:} is used for storing fine granular policies required for achieving end to end security in intra and inter domain communications. PRE stores security policies for all physical and virtual resources in a given domain, security profiles for each user and security policies for intra and inter domain communication. The security profile of a user contains fine granular information on the different devices and service specific requirements of that user. Then extraction component of PRE extracts the relevant fields from the repository PR which match with the attributes in the request and forwards them to SFEC. \\  
 \textbf{SFR:} stores the different security functions needed for the provision of 5G security services. In our architecture, SFR contains the following security functions: Trust Validation Function (TVF), Infrastructure Monitoring Function (IMF), Network Slice Access Function (NSAF), Flow Validation Function (FVF), Key Generation Function (KGF) and Flow Security Function (FSF). We will describe below how each of these functions are used in our architecture for mitigating  different attacks and achieving securing communications in 5G networks. \\
 \textbf{SFEC:} is used for enforcing the security policies on the various 5G infrastructure components via dynamic deployment of  the required security functions. In our architecture, the enforced polices are fine-grained such as flow specific, device specific, service specific, user specific, slice specific and domain specific functions. Network slice manager (in our case, SDN Controller) invokes these functions for securing the virtual/physical infrastructure from attacks. In addition, the users can also invoke these functions as part of security-as-a-service.\\
\textbf{ALC:} is used for maintaining a log of all activities and transactions of the SMA. 
\vspace{0.1cm}
\noindent Let us now consider the Security Function Repository (SFR) in our architecture.\\
\textbf {TVF}: The TVF uses TPM-based remote attestation to validate, at boot time, the state of the 5G infrastructure component, before VNF deployment. The attestation process involves measuring the state of the hardware and software parts of the infrastructure component in the form of hash values, and storing a signed version of these values in the form of trusted certificates. In principle, these trusted expected hash values can be created at any time. Most commonly this is done at boot time. Then, when a network slice makes an attestation request, the trusted platform provides an attestation report, which includes the measured hash values at the time of request.  The basic idea is that if there is a match between the measured hash values (at the time of request) and the expected trusted hash values (e.g. at boot time), then this indicates that the components are in a known and trusted state for deployment of VNFs. \\
In our architecture, the TVF on the SMA requests an attestation report from a physical or virtual infrastructure components and validates the report. The SMA will deploy security functions on the infrastructure component only if there is a match between the measured and expected hash values. Otherwise, the infrastructure component is considered to be compromised and an alert is raised to the SDN Controller administrator.\\
\textbf{IMF:} The TPM remote attestation typically only validates the state of the infrastructure component at boot time. However, the infrastructure components can be vulnerable to attacks during run time. IMF is used in the validation of the state of the infrastructure components at run time to detect attacks. For instance, in our architecture, we have developed functions for validating the flow rules enforced in the switches at run time. In this case, the IMF requests the network switch to report the set of current flow rules being enforced at the switch. The IMF then uses the trusted logs in the ALC to generate a trusted report on the expected state of the switch. The IMF compares the trusted report with the switch report to detect attacks, if there is any variation between them. \\
\textbf{NSAF:} The NSAF is deployed in the infrastructure component connected to the user devices (e.g. gNodeB). This function is used to ensure that only authorised devices are able to access the services in specific network slices. It also enforces decisions taken by Flow Validation Function (FVF), for instance, dropping requests from authorised devices which are found to be malicious by the FVF. An important advantage of the NSAF is that attacks from malicious devices are prevented closest to the source of attack, at the entry of 5G network.\\
\textbf{FVF:} This function enables network slice managers to monitor the usage of resources by authorised devices within the slices. FVF uses signature based and anomaly based attack detection mechanisms using machine learning techniques for attack detection within the slices.
If a flow from any device is found to be malicious, then it is terminated and an alert raised to SMA. The SMA dynamically isolates the flows from the malicious devices at the slice entry point by invoking NSAF. FVF can be invoked in the access network or the core or edge network depending on the latency requirements for the services in the specific slice. However, as FVF detects the attacks by monitoring the flows at the slice level, if fine-grained attack detection is needed, then users have the choice to select device specific security functions as part of security-as-a-service. \\
\textbf {KGF:} KGF is used to generate symmetric keys in the provision of security-as-a-service for securing communications of IoT applications or legacy devices in critical infrastructures. If the users have opted for securing communications, the SMA will invoke KGF to generate symmetric key for securing the flows and distribute the key to the communicating infrastructure components. \\
\textbf {FSF:} The FSF is invoked for securing communications between the end devices. It uses the symmetric key generated by the KGF for encrypting the flow at the component that is connected to the source device and decrypt the flow at the component that is connected to the destination device. Note that the flows are validated by the NSAF and FVF before they are encrypted by FSF. If the flow security policy mandates secure communication between the end devices, the SMA will use KGF to generate the required symmetric key for securing the flows and distribute the key to the relevant infrastructure components. \\
Industrial control infrastructures that were historically implemented as physically separate systems are being increasingly connected to the Internet for their maintenance/cost minimisation. Often such infrastructures have legacy systems which may not be able to support any security functionality. In such cases, the FSF can be used for securing communications between the end systems.\\ 
\textbf {Device Specific Security Function:} Users use this function to enforce policies that are specific to their devices. For instance, users can specify fine granular fingerprints for their devices and white or black lists for each of their devices.

\vspace{-4mm}

\subsection{Security Architecture - Operation}

\noindent Assume that a user has multiple devices registered with the SMA.  
The Security Profile of the user contains user’s personal attributes such as username and ID, role, and organization, and the user's service contract. The service contract specifies the attributes of the devices belonging to a user; these include the contractID, type of device, the number of each type of device, services that can be accessed by the registered user's device and the category this service belongs to (e.g. service X belongs to healthcare category) and the security requirements for the different services (such as confidentiality, integrity, authentication and accountability). A user is allowed to have multiple contracts. For instance, a user in his role as an employee of an organization X can have a service contract, contractID, whereas in a different role can have another contract. We have also defined a role called Personal-Role is defined for each user, which enables the user to have a service contract as an individual.
\noindent PR has policies associated with the users. For example, a simple representation of a policy rule is of the form \\
PRi= $<$RequestID; DeviceID; User{ID, name, ID, Role, Organization};{Services}; {Actions}$>$,
where i is the Policy Rule Identifier.\\
\noindent This rule defines the set of $<$Actions$>$ for a request from a user with specific attributes (such as UserID) for {Services} using a device DeviceID. \\
\noindent SFEC is used to establish secure communication flows and to monitor the ongoing communications for detecting different types of attacks based on the user's specific service contract.SFEC provides the main interface for interactions with the NSSF in the 5G core.\\
\textbf{Step 1:} Assume that a user has been registered with her devices UE1 and UE2. In this paper, we will not consider the authentication protocols that have been used to authenticate the user and the information during the registration process. This paper assumes the use of standard authentication protocols and that they are secure. Further, let us assume that UE1 is to be used to access a healthcare service whereas UE2 is to be used for accessing a financial service. These specifications have been defined as part of the policy specifications in the PRE as part of the registration process. \\
\noindent Now let us assume that UE1 is switched on to access a healthcare service.  UE1 now finds a 5G cell and sets up a radio connection with a gNodeB and generates a new flow request to gNodeB. The gNodeB has no information related to this specific flow from this user. \\
\textbf{Step 2:} gNodeB then forwards the request to Network Slice Selection Function (NSSF), which then sends the request to the SMA. We have abstracted several flows in the simplified diagram in Figure \ref{fig:secarc}. Normally the gNodeB selects the AMF to determine the network slices subscribed to the device which then contacts NSSF.
\vspace{-0.2cm}
\begin{itemize}
\item The 5G gNodeB selects an AMF that supports the network slices requested by the device. The 5G gNodeB becomes aware of the capabilities of the AMFs through the N2 interface.
\item The AMF learns about the network slices subscribed by the device from the UDM (after authentication – provided by UDM). \\ 
\noindent The AMF determines the allowed network slices for the device in the current registration.
\item If the AMF itself does not support all the slices requested by the device, then it seeks help from the NSSF to choose another suitable AMF. In such a case, the NSSF provides one or more allowed network slices for the device and works with the NRF to determine the set of AMFs that can provide these network slices.
\end{itemize}
\textbf{Step 3:} The NSSF then interfaces with the SMA through SFEC. The SFEC uses the PRE to check the policy rules and information associated with the user and his devices.
\vspace{-0.1cm}
\begin{itemize}
    \item PRE extracts the relevant fields from the user security profiles in the repository and forwards them to the SFEC. Here, there are different design choices possible. In our architecture, we pass the information on all network slices that the devices belonging to this particular user are subscribed to. Hence in this example, it will pass the network slices related to both the devices UE1 and UE2 of this user. This will include all the services that can be accessed using these two devices and the security requirements associated with all these services. Furthermore, during this step, logging of various actions carried out by the PRE are done by the ALC.
    \item SFEC then composes dynamically the Network Security Functions (NSF). This contains instances of different security functions from the repository SFR required based on the information from PRE in Step 3. In this example, it contains instances of two functions NSAF and FVF as shown in Fig 4. 
    \end{itemize}
\textbf{Step 4:} The SFEC then deploys the Network Security Functions (NSF) at the gNoneB via the NSSF through the standard 5G interfaces. A deployed function consists of VNF instances and associated resources (e.g. compute, storage, and networking resources). Furthermore, as in the previous step 3, again logging of various actions carried out by the SFEC are done by the ALC.\\
\textbf{Step 5:} First the Network Security Access Function (NSAF) instance is invoked at NSF to determine whether the user request using UE1 is allowed to access the services in the slice. This ensures that only the authorised devices and users are allowed to access the services in the specific network slice. In our design, we have allowed an unregistered device and user to be able to access a generic network slice (similar to a guest account) enabling limited communications over 5G network. An user can use this generic network slice to register his/her device following standard authentication protocols as mentioned above. Alternatively, the user can register his/her devices via out-of-band channel.  \\
Then the FVF instance is invoked at the network slice accessed by the user device to monitor the usage of resources by authorised devices within the network slice. As mentioned earlier, the FVF uses signature and anomaly based detection for enforcing flow based security policies. Note in this particular case, the FVF instance is being deployed at gNodeB. However, as shown in Figure~\ref{fig:accrnet} (see Appendix), FVF can be invoked in the access network or core network or edge network depending on the latency requirements for the services in the specific slice \cite{parvez2018survey}, \cite{ravindran20175g}, \cite{katsaros2017cache}. The design choice as to where FVF is deployed is dependent on the particular service and its provision. If for instance, the resource that needs to be monitored is a server in the cloud which provides the service, then it is more appropriate for FVF instance to be deployed close to the place where the service that is to be monitored is being run.\\
\textbf{Step 6:} Then UE1 sets up a connection in the healthcare slice to the Internet, enabling data transfer to the server providing the healthcare service. This establishes bi-directional end-to-end secure communications are established between UE1 and the health service provider with appropriate 5G quality of service. In our case, from the point of view of security, this corresponds to appropriate security mechanisms such as confidentiality and integrity (based on the information in PRE), implemented using VNFs, ensuring that the flows are protected.\\
\textbf{Step 7:} Now consider another device UE2 belonging to the same user initiates a request to access a different service, a financial service slice in 5G network. The NSF in gNodeB already has information related to this flow request for UE2. Recall the design choice made in our architecture is that  all information related to all network slices the devices of a given user are subscribed to (and the services that can be accessed by these devices and their associated security requirements) extracted from the PRE, are deployed in NSF by the SFEC.  Hence NSAF and FVF are automatically applied for the device UE2 as well.\\
\textbf{Step 8:} As in Step 6, bidirectional end-to-end secure communications can now be established between the device UE2 and the the financial service provider with appropriate security mechanisms applied to the communication flows.
\vspace{-3mm}
\subsection{Example Scenario and Mitigation of Attacks}
\vspace{-2mm}
\textbf{Attack Scenario 1:} Consider the attack case where the attacker has compromised one of the user’s devices and it is used to generate DoS attack on different network slices. This could happen for instance if one of the smart home devices such as a home printer is compromised by Mirai botnet and is used to flood the healthcare service slice in 5G network.\\ 
Using our security architecture, only the devices that are authorised to have the capability to access the healthcare service slice will be able to do so. So the flow from the home printer will be first checked by the NSAF, which will detect that the flow is not authorised to access healthcare slice and hence will drop the flow.\\
\textbf{Attack Scenario 2:} Now consider the situation where a wearable healthcare sensor device, which is authorised to access the healthcare slice generates the flooding attack. In this case, NSAF will permit the flow from the wearable sensor to the healthcare slice. However, the signature or anomaly-based detection in FVF detects the suspicious flows from wearable healthcare sensor and raises an alert to SMA. SMA then dynamically configures the NSAF to drop the flows from malicious wearable sensor at NSAF. Note that in this case, it is more efficient for the NSAF to be implemented at the entry point of the slice at NSF rather in the access network or core network, Hence our approach is able to deal with such attacks in an efficient manner.\\
\textbf{Attack Scenario 3:}  Now consider the situation where the server providing the service to a user device (such as the healthcare wearable sensor) has been infected with malware. The SMA deploys the Trust Validation Function (TVF) at the server to validate the state of the server before the server is allowed to provide the service to the user device. In our architecture, this is achieved at by specifying policy rules in PRE which require state validation of the server prior to service installation and provision. This in turn mitigates attacks that lead to provision of malicious services. The state validation of the server at run time is achieved using the Infrastructure Monitoring Function (IMF) at runtime.\\ 
\textbf{Attack Scenario 4:} Now consider the case where a user’s device UE accesses a service via radio access network node gNodeB and then moves to a different radio access networks gNodeB’ belonging to the same 5G network.  The initial network flow from the device UE was analysed to determine whether it was authorised or not by the NSSF. Because both the radio networks belong to the same Evolved Packet Core (EPC) and under the same SMA, the established authorisations are handed over by the NSSF in gNodeB to gNodeB’.  When the UE moves to another EPC, handover between different SMA in different domains in 5G networks need to happen. We will not consider this inter-domain mobility scenario and the extensions to our security architecture in this paper. These are currently being developed and we will address them in a separate paper.
\vspace{-0.4cm}
\section{Implementation and Analysis} \label{sec:imp}
\vspace{-0.3cm}
\noindent We have developed a solution implementing our security architecture using a SDN Controller, in which we have combined several authorities of the 5G NFV architecture performing different functionalities into a single combined authority. We have used a ONOS SDN Controller \cite{berde14} and our simulation experiments have been carried out using Mininet. In this section, we present the implementation and analysis of our security architecture.
\vspace{-0.4cm}
\subsection{Prototype Implementation}
\begin{figure}
\centering
\includegraphics[scale=.45]{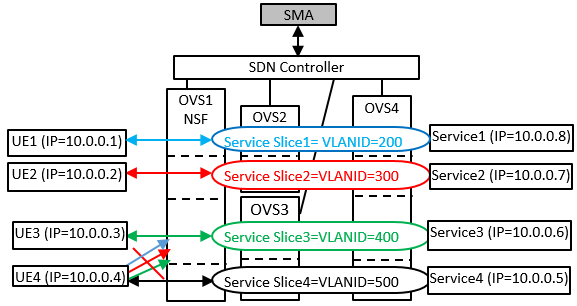}
\caption{Implementation Scenario}
\label{fig:5gsim}
\end{figure}

 \vspace{-0.2cm}
\noindent The SMA 
has been implemented as an application of the ONOS SDN Controller \cite{berde14} Version 1.6.0 using Java and Integrated Development Environment for securing 5G networks. 
We have used the ONOS application development interface to implement the SMA. We have added extra modules to ONOS implementing the various SMA functions. 
We have created specific Mininet scripts to create the topology and simulations have been carried using a varying number of hosts and switches.
\noindent For easy presentation, we present the operation of our architecture using simple network topology shown in Figure~\ref{fig:5gsim}, with four network slices represented in blue, red, green and black that are connected to specific services. 
The slices are implemented using VLANs and the slices are distributed among multiple paths as shown in the Figure~\ref{fig:5gsim}.   
\vspace{-5mm}

\begin{figure}
\centering
\includegraphics[scale=.28]{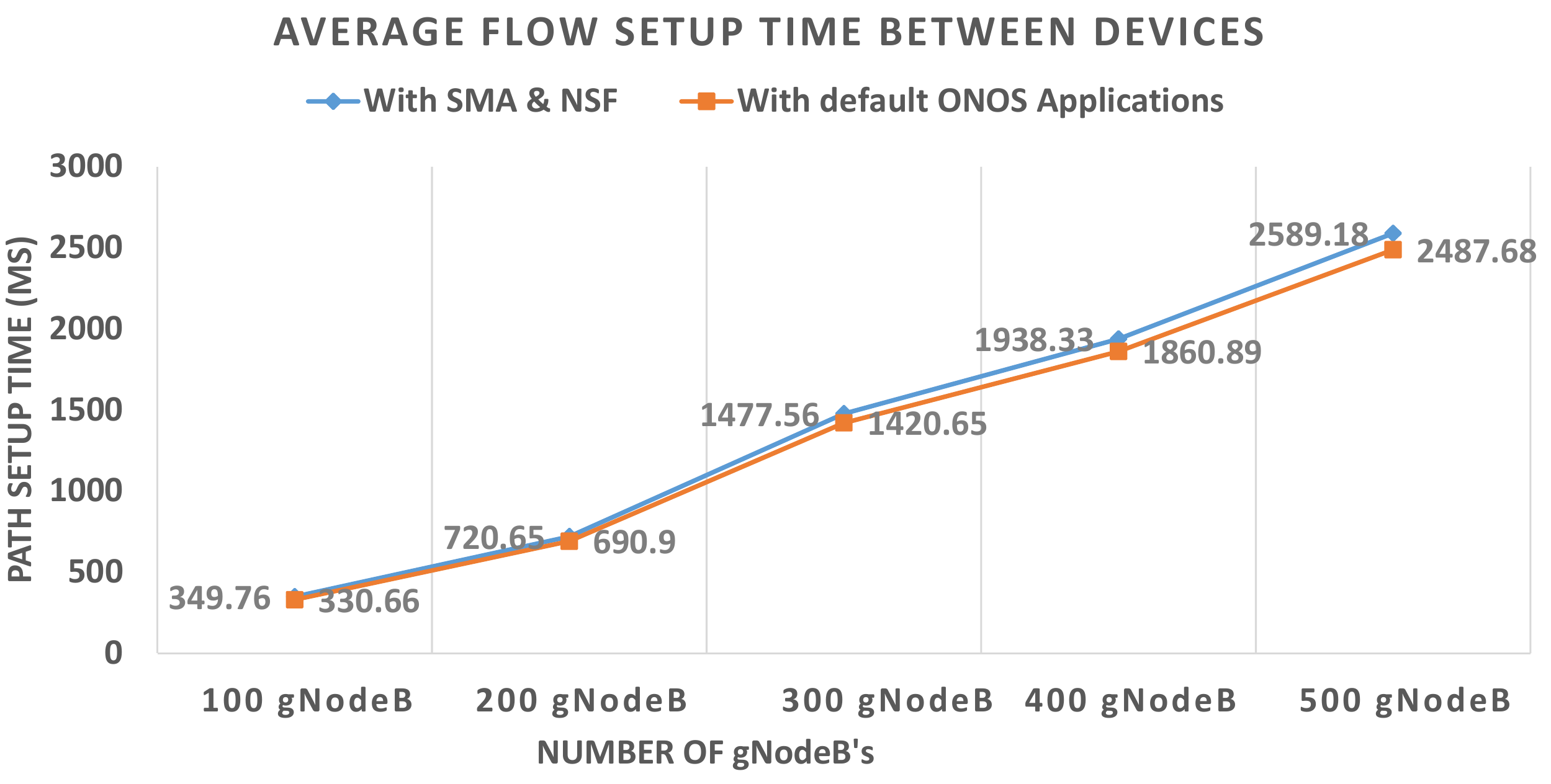}
\caption{Average flow setup time}
\label{fig:nrtime}
\end{figure}

In the scenario shown in Figure~\ref{fig:5gsim}, a user has 4 devices(UE'S), where each UE is registered to access a specific slices (VLAN's) and specific services in the 5G network. The IP address of the UE's and services and the VLANID of the slices are also shown in the Figure~\ref{fig:5gsim}. The SMA creates and assigns flows from different user devices to these VLANs. The mininet topology for this scenario is shown in Figure~\ref{fig:vtopo2} (see appendix). The slice access policies are stored in a JSON repository and SMA uses these  policies to transfer the user flows through different slices. For user and device details, we have created a user to device relational database. Here, each user has multiple devices and each device has associated with it different policy rules and actions. The action consists of access to specific slices and related security functions. The policies can apply to users, devices, flows or services (see Appendix: Figure \ref{fig:db} shows the SMA repository schema and a sample policy for UE1 is presented in the Listing 1). To update, modify and read the stored JSON policies, we have used the JAVA parser module. We have developed additional modifications to the Mininet scripts to enforce the flow isolation based on the Controller response.\\
\noindent When OVS1 (representing gNodeB) receives the first flow from a user device to access a specific slice, it forwards the packet header to the Controller, which is then forwarded to SMA. The SMA analyses the header to determine the specific policies related to the user device and dynamically deploys the NSFs to validate the flows from different user devices against attacks. \\
\noindent The security functions for virtual OpenFlow switches are implemented in Xen VMM. 
We have used OVS version 2.5.0 on Xen 4.6 VMM with Dom0 running Ubuntu 15.10. 
The SMA enables the NSF( ) security functions in the Dom0. The NSAF( ) captures the VM traffic from the virtual interface and analyses the flow from the virtual machine. If the flow is accessing a service slice that is not registered for the UE (virtual machine), then NSAF drops the packet and raises an alert to SMA. If the flow is accessing a registered service slice, then the flow is validated using FVF( ) using signature based detection and anomaly based detection, and then forwarded to the specific service slice. If the flow matches any of the attack signatures or found to be suspicious by the anomaly detection module, then it is dropped and an alert is raised to the SMA.  \\
There have been prior studies on the performance of different Controllers (e.g. \cite{shalimov13}). However, with our security solution, what we found is that the main overhead has been due to the enforcement of security policies with the SMA at the switches using the NSF(). Hence we have been primarily concerned with this performance issues such as variation in end to end path setup time and throughput with SMA and NSF, increase in latency with number of rules for traffic validation, CPU overhead and  memory overhead. We have also extended the LINC-Switch \cite{lisw} for dynamic encryption/decryption of flows using the  \textit{crypto} module. We have used AES algorithm with key size 128 bits and analysed the end to end delays for dynamic encryption of the flows at the ingress node and decryption of the flows at the egress node. Due to space restriction we are only presenting some of the results in this paper. \\
The average flow setup time between UE's and the services increases with the varying number of gNodeB's (Figure \ref{fig:nrtime}). With ONOS default applications running, the average flow setup time for 100 gNodeB's is $330.66ms$ and it steadily increases to $2487.68ms$ with 500 gNodeB's. On the other hand, with SMA and NSF running over the controller causes some delay in flow setup time. In this case, with 100 gNodeB's the average flow setup time is $349.76ms$ and it increases to $2589.18ms$ with 500 gNodeB's. 
\vspace{-3mm}
\subsection{Attack Detection using FVF( )}
\begin{figure}
\centering
\includegraphics[scale=.4]{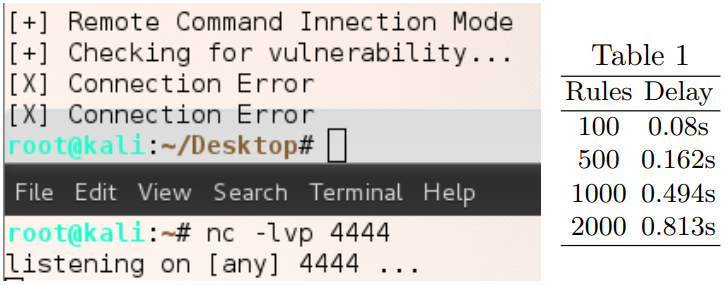}
\caption{Attack Prevention at source NSF}
\label{fig:pencon}
\end{figure}

\vspace{-2mm}
\noindent Consider a service slice that is related to critical services such as in a SCADA system. Often critical infrastructure operators are hesitant to apply the patches to their vulnerable services, as this may require the systems to be taken offline, which may not be possible. Furthermore, sometimes operators are also worried that patching itself can break the SCADA application. However, as long as the systems remain unpatched, they are vulnerable to attacks. FVF() helps to protect such vulnerable systems by preventing the malicious traffic at the gNodeB (switch) that is near to the source of attack. In this case, we have used a Kali Linux virtual machine as the malicious client (UE1) to launch a Shellshock attack on an unpatched Web Server (Service1) and attack signatures are applied at NSF in OVS1 using FVF( ). 
As shown in Figure~\ref{fig:pencon}, the attack was not successful and the malicious UE1 is isolated from the network. There is latency with the use of signatures in flow validation. Table 1 presents the average latency results for 10 runs for a varying number of signatures used for flow validation.

\vspace{-4mm}

\noindent We have used the ITOC dataset for evaluating the anomaly detection functionality of the FVF( ). 
Table 2 shows the results for anomaly detection using different machine learning techniques for the ITOC dataset. 
We have also compared the performance of Naive Bayes (NB), Decision Tree C 4.5 (DT) and Random Forest (RF) classifiers with Chi Square and with an ensemble of feature selection methods Recursive Feature Elimination (RFE) and Chi Square . As shown in Table 2, RF when executed with an ensemble of RFE and Chi Square achieved much better accuracy of 98.888\%. Figure~\ref{roc} shows the Receiver Operating Characteristics (ROC) curve plotted against the TPR and FPR for the machine learning techniques executed with an ensemble of RFE and Chi Square.
\vspace{-5mm}
\begin{table}[]
\centering
\label{tab:titoc}
\footnotesize
\begin{tabular}{|c|c|c|c|c|c|}
\hline
\textbf{\begin{tabular}[c]{@{}c@{}}Classifier+ Feature selec.\end{tabular}} & \textbf{Accuracy} & \textbf{TPR} & \textbf{TNR} & \textbf{FNR} & \textbf{FPR} \\ \hline
NB (Chi Square) & 68.760 & 51.100 & 81.544 & 48.89 & 18.455 \\ \hline
DT (Chi Square) & 92.740 & 88.54 & 97.7914 & 11.458 & 4.208 \\ \hline
RF (Chi Square) & 97.070 & 97.052 & 97.085 & 2.947 & 2.914 \\ \hline
\begin{tabular}[c]{@{}c@{}}NB (RFE+ Chi Square)\end{tabular} & 68.723 & 51.112 & 81.460 & 48.887 & 18.539 \\ \hline
\begin{tabular}[c]{@{}c@{}}DT(RFE+ Chi Square)\end{tabular} & 94.659 & 89.800 & 98.17 & 10.199 & 1.8266 \\ \hline
\begin{tabular}[c]{@{}c@{}}RF (RFE+ Chi Square)\end{tabular} & 98.888 & 99.182 & 98.482 & 0.8176 & 1.517 \\ \hline
\end{tabular}
\end{table}
\subsection{Attack Detection on Infrastructure Elements}
\noindent Let us now consider how SMA is used to detect attacks on the infrastructure components in the 5G network. \\
SMA makes use of TVF( ) to validate, at boot time state of the switches before configuring flow rules on the switches and uses IMF ( ) is used to detect the attacks on the switches at runtime. The IMF requests the network switch to report the set of current flow rules being enforced at the switch. The IMF then uses the trusted logs in the ALC to generate a trusted report on the expected state of the switch. The IMF compares the trusted report with the switch report to detect attacks, if there is any variation between them. We have used VIM diff tool for comparing the switch\_state\_report with the trusted report.The operation of the IMF () for a legitimate scenario is shown in Figure~\ref{flba}. 

Now consider that an attacker has initiated attack on the switch OVS1 to alter some of the flow rules. In this case, we used Kali Linux machine (UE1) to generate malicious flow\_mod message to insert an unauthorised flow rule in OVS1. 
\begin{figure*}[!ht]
    \centering
    \includegraphics[scale=.3]{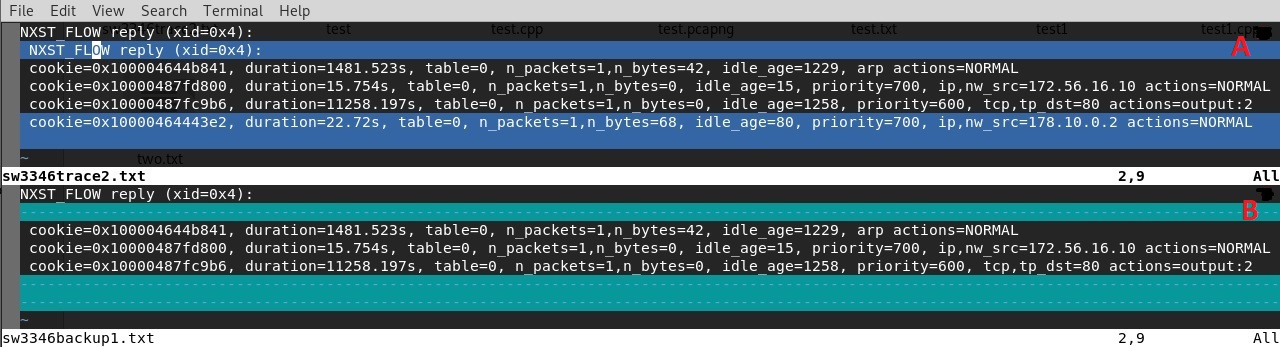}
    \caption{A) Switch Report; B) Trusted report}
	\label{aa}
\end{figure*}
At a later time, the IMF( ) in the SMA first sends a request to the switch OVS1 (ID=3346) to report the flow rules. The switch\_state\_report from switch OVS1 is shown in the top Window A in Figure~\ref{aa} which also includes the flow rule inserted by the attacker. Now, IMF( ) generates a trusted report on the expected state of the switch SW1 by querying the ALC database for switch ID=3346. The trusted report on the expected state of the switch is shown in the below Window B in Figure~\ref{flba}. The IMF( ) compares these reports using VIM diff tool. The IMF( ) detects this malicious flow rule when it compares the switch\_state\_report with the trusted report shown below in the Window B in the Figure~\ref{aa}. So it invokes the alert and restore component. The alert and restore component raises an alert to the SMA Security Administrator. 


\vspace{-0.4cm}

\section{Related Work} \label{sec:rel}
\vspace{-0.3cm}
\noindent In this section, we briefly discuss important related works that are relevant to our proposed security solution and architecture. \\
SDN and NFV are the key enabler technologies for 5G network \cite{yousaf2017nfv}, \cite{ordonez2017network}. For instance, techniques such as  \cite{ordoez2018creation}, \cite{han2019utility} have been proposed for provisioning slice as a service model to the tenants. The work in \cite{guija2018identity} proposed SDN and NFV based authentication and authorisation framework for  Identity  and  Access  control  of micro-services in  5G  platforms  for  services  virtualization, orchestration and management. Our architecture assumes usage of standard authentication protocols in 5G networks. 
The work in \cite{siddiqui2016policy} makes use of SDN and NFV for developing a security architecture framework with components such as policy-based security management system, Service monitoring \& analytics systems and Virtual security functions to achieve desired security functionality. The Security Policy Manager is used for structuring the security requirements and deploying the required virtual security functions. The Service Monitoring \& Analytics component deployed in the orchestration layer is responsible for performing metrics and
notifications acquisition from virtual and physical infrastructure components. However it does not present detail discussion on how the architecture is used to deal with specific attacks. We have presented a detail discussion on our security architecture components and how it can be used to deal with the specific attacks in 5G networks.\\
Although there are significant benefits with SDN and NFV, security in SDN and NFV is still at its infancy and security issues in these technologies \cite{kreutz13} \cite{schehlmann14}, \cite{hong15}, \cite{dargahi2017survey}, \cite{lee2017delta}, \cite{dacier2017} also impact the security in 5G networks \cite{ting2019guidelines}, \cite{barakabitze20205g}, \cite{badmus2020identifying}, \cite{ahmad20175g}. Hence we need to consider the security techniques that have been proposed to enhance the security in these technologies. Porras et al. \cite{porras12} proposed security enforcement kernel for securing the Controllers from northbound applications. It uses role based access control to prevent northbound applications from making unauthorised changes to the database in the Controller. Shin et al. \cite{shin2014rosemary} proposed 
techniques for sandboxing the network applications to prevent common failures of applications from halting the SDN Controller operation. Similar techniques can be used with our security solution to protect the SMA from malicious SDN Controller applications and improve the resilience of the Controllers.\\
There is some work on detecting the attacks on infrastructure components such as switches. Jacquin et al. ~\cite{jacquin15towards} proposed using TPM attestation to validate the boot time state of the switches. However there can be attacks on the switches during runtime which have not been addressed.  Zhang \textit{et al.} in \cite{zhang17towards} 
proposed rule enforcement verification to detect rule modification attacks in OpenFlow switches. REV uses recursive message authentication code to verify whether the OpenFlow switches have enforced the flow rules. Our architecture makes use of TVF, IMF and ALC in the SMA to detect attacks on switches. TVF is used to to validate the boot time of switches. IMF and ALC are used to generate trusted report on the expected state of switches to validate the flow rules enforced and detect attacks during runtime. \\
In the current networks, denial of service of attacks are often detected or prevented at the victim end and traceback techniques \cite{savage01}, \cite{stone00}, \cite{mahajan02} are used to determine the approximate source of the attack. Applications in 5G networks have stringent latency requirements and such attacks can have severe impact in 5G networks. Our architecture makes use of NSF()  to validate the traffic generated by the end nodes and drop the attack traffic at the source end.
\vspace{-0.45cm}
\section{Concluding Remarks} \label{sec:con}
\vspace{-0.3cm}
\noindent In this paper we have presented a security architecture and its facilities and mechanisms for securing 5G networks against specific security attacks. We have developed and implemented a simplified version of the security architecture using SDN and NFV technologies. We have demonstrated that the proposed architecture is able to counteract attacks from malicious devices accessing 5G services in an unauthorised manner or generating flooding and other denial of service attacks. We have also provided mechanisms to ensure that the network functions are deployed in trusted infrastructures thereby ensuring that 5G services are not provided by malicious infrastructure components. Our security architecture also enables security-as-a-service in 5G networks, thereby allowing the provision of security services in the form of network slices to devices and legacy systems which are otherwise constrained and lack security functionalities. We have implemented the proposed architecture and carried out experiments and discussed results demonstrating how it can be used to counteract different types of attacks. We are currently extending the architecture to counteract attacks in a multiple 5G network operators' environment. Furthermore, as part of this research work, we are developing security services enabling dynamic establishment of trusted end to end 5G services over multiple domain networked virtualised infrastructures.


\bibliographystyle{splncs04} 
\bibliography{ref}

\begin{figure}[!ht]
\centering
\includegraphics[scale=.3]{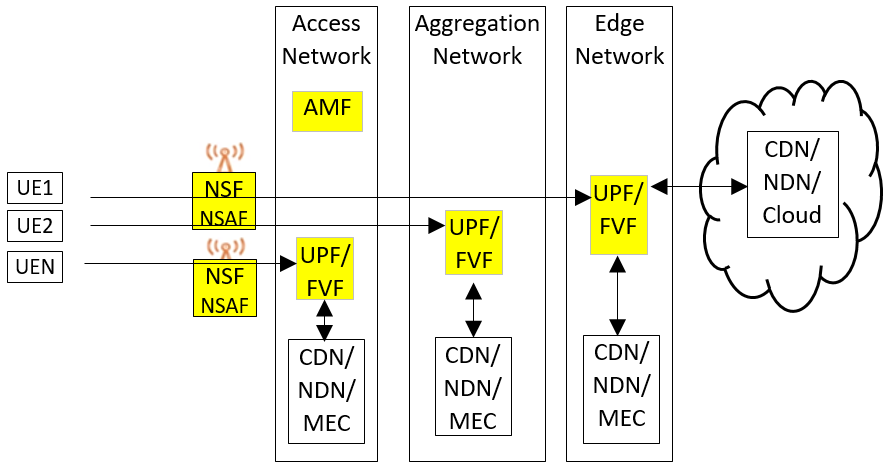}
\caption{Network Scenario}
\label{fig:accrnet}
\end{figure}

\begin{lstlisting}[caption={A Sample Policy}, label={list:pe}] [!ht]
{
  "id": "02",
  "hostip": "10.0.0.1",
  "hostmac": "00:09:00:AA",
  "destip": "10.0.0.8",
  "dstmac": "00:09:00:BB",
  "flowid": "78b34x",
  "actions": [ 
    "Service": "Service1",
    "Slice-id":"VLAN200",
             ],
  },
}
\end{lstlisting}

\begin{figure*}[!ht]
\centering
\includegraphics[scale=0.45]{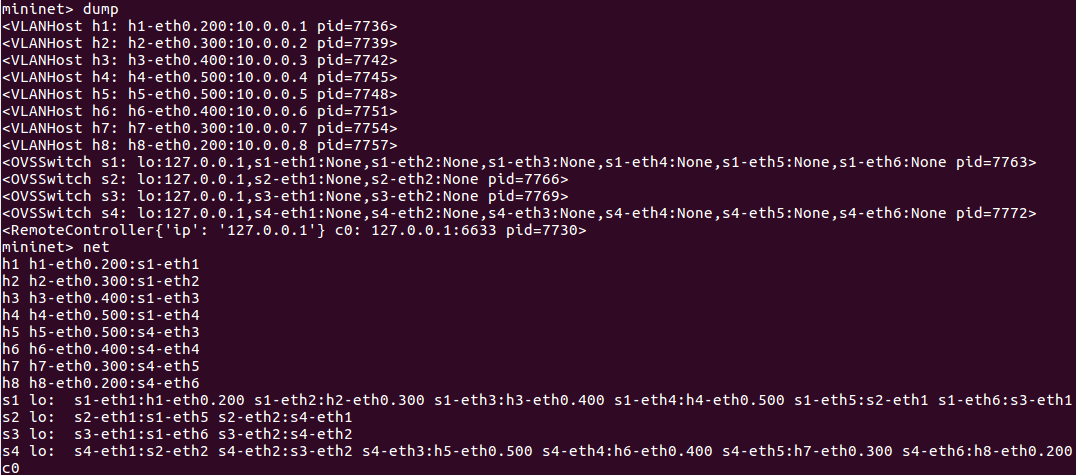}
\caption{Mininet Network View}
\label{fig:vtopo2}
\end{figure*}

\begin{figure}[!ht]
\centering
\includegraphics[scale=.4]{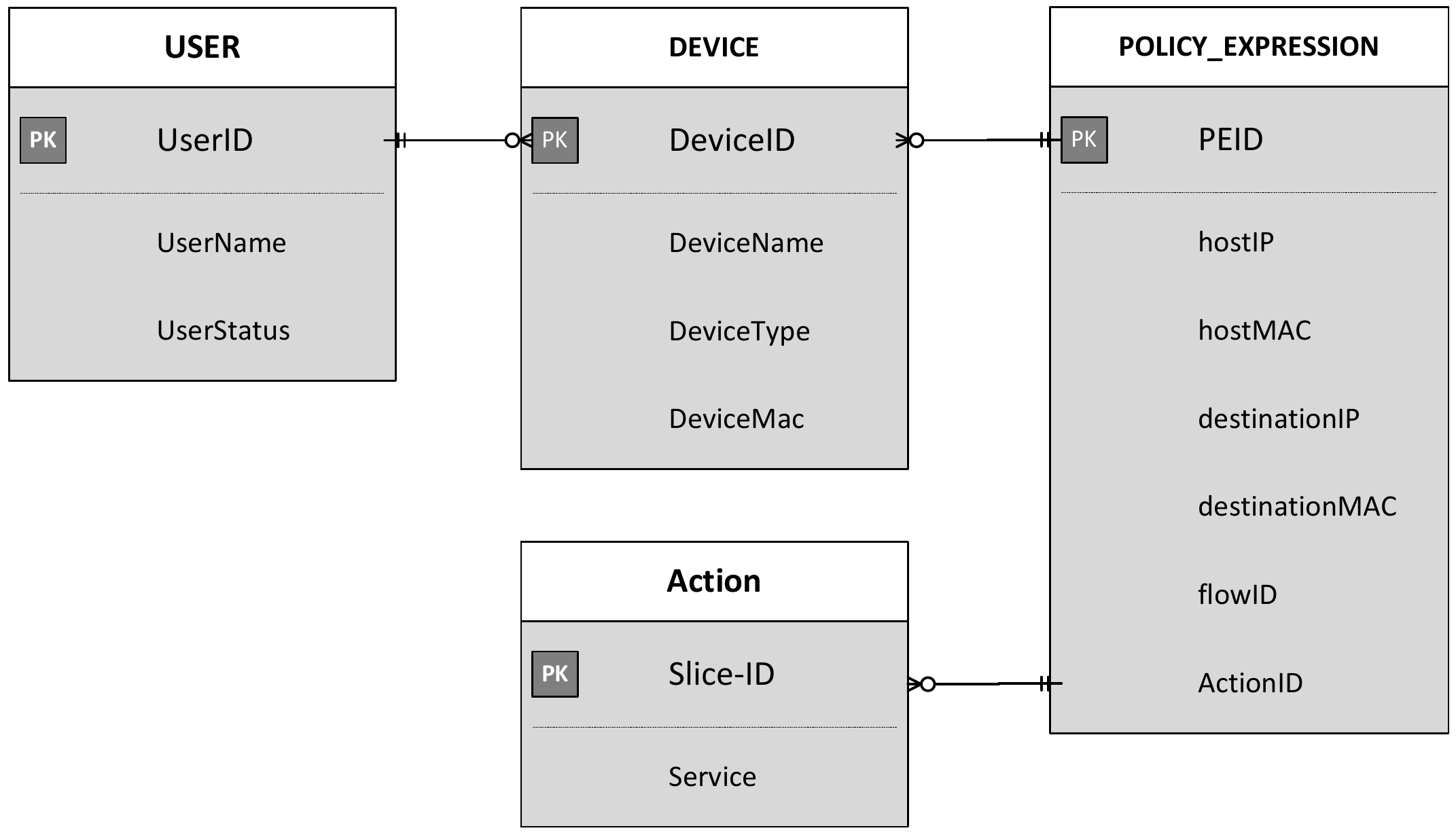}
\caption{ACA Repository Schema}
\label{fig:db}
\end{figure}

\lstset{
    string=[s]{"}{"},
    stringstyle=\color{blue},
    comment=[l]{:},
    commentstyle=\color{black},
}

\begin{figure}
\centering
\includegraphics[scale=0.6]{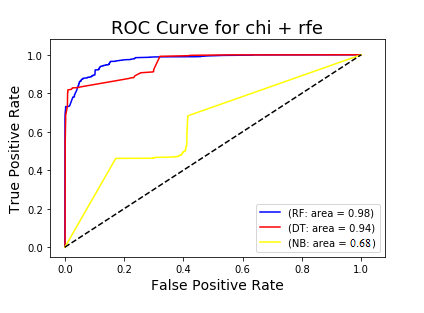}
\caption{ROC Curves}
\label{roc}
\end{figure}

\begin{figure*}[!ht]
    \centering
    \includegraphics[scale=.35]{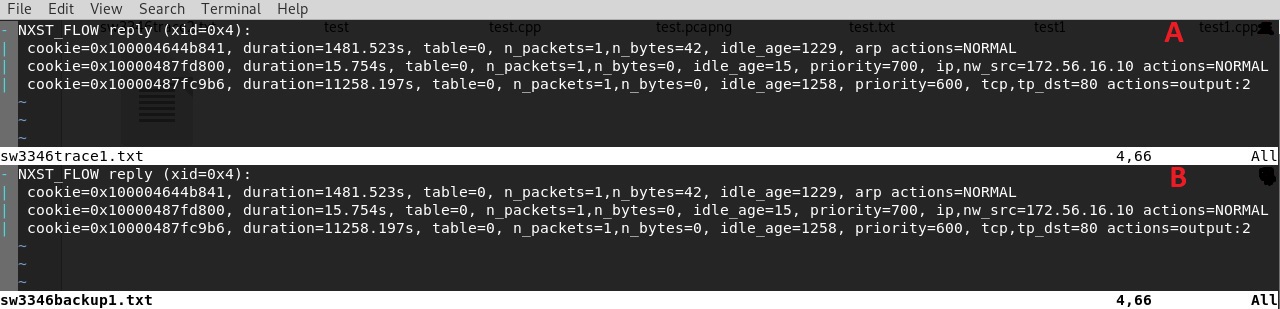}
    \caption{A) Switch Report; B) Trusted report}
	\label{flba}
\end{figure*}


\end{document}